\documentstyle[emulateapj,psfig,apjfonts]{article}
\def\gs{\mathrel{\raise0.35ex\hbox{$\scriptstyle >$}\kern-0.6em 
\lower0.40ex\hbox{{$\scriptstyle \sim$}}}}
\def\ls{\mathrel{\raise0.35ex\hbox{$\scriptstyle <$}\kern-0.6em 
\lower0.40ex\hbox{{$\scriptstyle \sim$}}}}
\submitted{Received: July 01, 1999; Accepted August 20, 1999}

\lefthead{Smail et al.}
\righthead{Radio Observations of Submm Galaxies}

\begin{document}

\title{Radio Constraints on the Identifications and Redshifts of Submm Galaxies}

\author{Ian Smail,$\!$\altaffilmark{1,2} R.\,J.\ Ivison,$\!$\altaffilmark{3,4}
F.\,N.\ Owen,$\!$\altaffilmark{5} A.\,W.\ Blain\altaffilmark{6} \& J.-P.\ Kneib\altaffilmark{7}
}
\affil{\small 1) Department of Physics, University of Durham, South Road, 
Durham DH1 3LE, UK}
\affil{\small 3) Department of Physics \& Astronomy, University College London, 
        Gower Street, London WC1E 6BT, UK}
\affil{\small 5) NRAO, P.O.\ Box 0, 1003 Lopezville Road, Socorro, NM 87801}
\affil{\small 6) Cavendish Laboratory, Madingley Road, Cambridge
CB3 OHE, UK}
\affil{\small 7) Observatoire Midi-Pyr\'en\'ees, CNRS-UMR5572, 14 Avenue E.\ Belin, 31400 Toulouse, France}

\setcounter{footnote}{5}

\altaffiltext{2}{Royal Society University Research Fellow.}
\altaffiltext{4}{PPARC Advanced Fellow.}

\begin{abstract}
We present radio maps from the Very Large Array (VLA) for 16
sources detected in a sub-millimeter (submm) survey of the distant
Universe. Our deep VLA 1.4-GHz maps allow us to identify radio
counterparts or place stringent limits ($\ls 20\mu$Jy in the source
plane) on the radio flux of the submm sources.  We compare the spectral
indices of our sources between 850\,$\mu$m and 1.4\,GHz to empirical
and theoretical models for distant starburst galaxies and active
galactic nuclei (AGN) as a function of redshift.  In this way we can
derive redshift limits for the submm sources, even in the absence of an
optical or near-infrared counterpart.  We conclude that the submm
population brighter than $\sim 1$\,mJy has a median redshift of {\it at
least} $<\! z\! >\sim 2$, more probably $<\! z\! >\sim 2.5$--3, with
almost all galaxies at $z\gg 1$.  This estimate is a strong lower limit
as both misidentification of the radio counterparts and non-thermal
emission from an AGN will bias our redshift estimates to lower
values.    The high median redshift means that the submm population, if
predominately powered by starbursts, contributes a substantial fraction
of the total star formation density at high redshifts.  A comparison of
the spectral index limits with spectroscopic redshifts for proposed
optical counterparts to individual submm galaxies suggests that 
half of the submm sources remain unidentified and thus their
counterparts must be fainter than $I\gs 24$.
\end{abstract}



%
%
%
\section{Introduction}

Faint submm sources are likely to be highly obscured starburst galaxies
and AGN, within which optical/UV radiation from massive stars or an active
nucleus is absorbed by dust and reradiated in the far-infrared.  The
dust emission peaks at $\lambda \sim 100\mu$m and thus long-wavelength
observations of distant dusty galaxies can benefit as this peak is
redshifted into their window.  At 850\,$\mu$m, this increase balances
the geometrical dimming at higher redshifts, resulting in a
constant flux density out to $z\sim 5$--10 and the opportunity to
select very high redshift galaxies.

%
%
\begin{figure*}[tbh]
\begin{center}
\psfig{file=f1.ps,width=7.0in,angle=270}
\end{center}
\vspace*{-4mm}
\noindent{
\scriptsize \addtolength{\baselineskip}{-3pt} 
Fig.~1. $15\times 15$\,arcsec VLA maps of the 16 submm sources for
which we have radio observations.  These are ordered as in Table~1 and
the maps span a range in sensitivity and resolutions (\S2.1).  The
panels are centered on the nominal submm position in each case and we
show the typical error boxes for the sources.  We mark the radio
counterpart by a $+$  where they are identified.  The panels have north
top and east left and correspond to $\gs 100 h^{-1}_{50}$\,kpc at
$z>1$.  The contour levels are at apparent 1.4-GHz flux densities of 3, 5, 10,
20, 50, 100$\times$ the map noise listed in \S2.1 (except for
SMM\,J21536+1741 and SMM\,J14010+0252 where they are 100$\times$ these
values). \hfil

\addtolength{\baselineskip}{3pt}
}\vspace*{-6mm}
\end{figure*}

A number of deep submm surveys have been published (Smail, Ivison \&
Blain 1997; Barger et al.\ 1998, 1999b; Hughes et al.\ 1998; Blain et
al.\ 1999a; Eales et al.\ 1999) providing counts which are in good
agreement at 850-$\mu$m flux densities above 2\,mJy (the confusion
limit of the blank-field surveys).  The surface density of submm
galaxies reaches $\sim 2$ per sq.\ arcmin by 1\,mJy (Blain et
al.\ 1999a). If these galaxies lie at $z\gs 1$, then they have
bolometric luminosities of $\gs 10^{12} L_\odot$ and they are the
distant analogs of the local ultraluminous infrared galaxy (ULIRG)
population.  However, the observed surface density of submm galaxies is
several orders of magnitude greater than that expected from the local
ULIRG population (Smail et al.\ 1997) indicating very substantial
evolution of these systems in the distant Universe.  The integrated
emission from this population can account for the bulk of the
extragalactic background detected at 850\,$\mu$m by {\it COBE}
(e.g.\ Fixsen et al.\ 1998), and hence confirms these galaxies as an
important source of radiation in the Universe (Blain et al.\ 1999a, 1999b).

To identify the era of obscured emission in the Universe, whether from
AGN or starbursts, we have to measure the redshifts of a complete
sample of submm galaxies.  Several groups have attempted this (Hughes
et al.\ 1998; Barger et al.\ 1999a; Lilly et al.\ 1999).  Hughes et
al.\ (1998) concluded that the bulk of the population is at $z\sim
2$--4, based on photometric redshift limits for the probable
counterparts of five submm sources in the {\it Hubble Deep Field} ({\it
HDF}, c.f.\ Richards 1999 and Downes et al.\ 1999).  Barger et
al.\ (1999a) undertook a spectroscopic survey of the same submm sample
analysed here and concluded that the median redshift was $<\! z\! >\sim
1.5$--2, with the bulk of the population having $z\sim 1$--3.  Finally,
Lilly et al.\ (1999) used archival spectroscopy and broad-band
photometry of submm sources from the Eales et al.\ (1999) survey to
claim that the population spans $z=0.1$--3, with a third at $z<1$.  The
differences between these studies are significant and important for our
understanding of the nature of submm galaxies.

It is very difficult to achieve high completeness in optical
spectroscopic surveys of submm galaxies (e.g.\ Barger et
al.\ 1999a) due to the very different behaviour of the K corrections for
distant galaxies between  submm and optical passbands.  However,
even a crude estimate of the median redshift of a {\it complete} sample
of submm galaxies would provide a powerful insight into the relative
dominance of obscured and unobscured emission at different epochs
(Blain et al.\ 1999b).

In a recent paper, Carilli \& Yun (1999, CY) demonstrated that using
the spectral index between the submm (850\,$\mu$m) and radio (1.4\,GHz)
wavebands, $\alpha^{850}_{1.4} = 0.42 \log_{10} (S_{850}/S_{1.4})$, it
was possible to obtain crude redshift limits for distant dusty
galaxies, irrespective of the nature of the emission mechanism, AGN or
starburst.  CY employed a number of theoretical and empirical spectral
energy distributions (SEDs) to investigate the range in
$\alpha^{850}_{1.4}$  for different assumed SEDs and showed that these
models adequately described the small sample of high-redshift galaxies
for which both radio and submm observations were available. As pointed
out by Blain (1999, B99), if we adopt lower dust temperatures for the
submm population than are seen in the local sources used in CY's
models, then the allowed range of redshifts is slightly lower for a
given value of $\alpha_{1.4}^{850}$.  Nevertheless, the modest scatter
between the models in CY suggests that this technique can provide
useful limits on the redshifts of submm galaxies in the absence of an
optical counterpart.
  
In this paper we apply the CY analysis to deep radio observations of
a complete sample of submm galaxies selected from the SCUBA Cluster
Lens Survey (Smail et al.\ 1998).  Our aim is to constrain the redshift
distribution of this population and in the process test the optical
identifications and spectroscopic redshifts from Smail et al.\ (1998)
and Barger et al.\ (1999a).  We present the observations and their analysis
in \S2, discuss our results in \S3 and give our main conclusions in \S4.

\section{Observations, Reduction and Analysis}

The 850-$\mu$m maps on which our survey is based were obtained using
the long-wavelength array of the Sub-millimeter Common-User Bolometer
Array (SCUBA, Holland et al.\ 1999) on the James Clerk Maxwell
Telescope (JCMT)\footnote{The JCMT is operated by the Joint Astronomy
Centre on behalf of the United Kingdom Particle Physics and Astronomy
Research Council, the Netherlands Organisation for Scientific Research,
and the National Research Council of Canada.}.  The details of the
observations, their reduction and analysis are given in Smail et
al.\ (1997, 1998) and Ivison et al.\ (1998).  Each field covers an area
of 5.2\,arcmin$^2$ with a typical 1$\sigma$ sensitivity of
$\sim$1.7\,mJy, giving a total survey area of 0.01\,deg$^2$.  The
median amplification by the  cluster lenses for background sources
detected in our fields is expected to be $2.5^{+5.0}_{-1.5}$ (Blain et
al.\ 1999a; Barger et al.\ 1999a), and so we have effectively surveyed
an area of about 15\,arcmin$^2$ in the source plane to an equivalent
$1\sigma$ sensitivity of 0.7\,mJy.  The follow-up of these submm
sources also benefits from the achromatic amplification, which boosts
the apparent brightness of counterparts in all other wavebands.

\subsection{Radio Observations}

All the radio maps used in this work were obtained with the
VLA\footnote{The VLA is run by NRAO and is operated by Associated
Universities Inc., under a cooperative agreement with the National
Science Foundation.} at 1.4\,GHz in A or B configuration, giving
effective resolutions of 1.5$''$ and 5$''$, respectively.  More details
of the reduction and analysis of these maps are given in the following
references (we list the VLA configuration and 1$\sigma$ map noise for
each cluster):  Morrison et al.\ (1999) for Cl\,0024$+$16
(B/15\,$\mu$Jy), A\,370 (B/10\,$\mu$Jy) and Cl\,0939$+$47
(B/9\,$\mu$Jy); Ivison et al.\ (1999) for A\,1835 (B/16\,$\mu$Jy) and
Ivison et al.\ (2000) for MS\,0440$+$02 (A/15\,$\mu$Jy) and
Cl\,2244$-$02 (A/17\,$\mu$Jy).  No deep radio map is available for
A\,2390, although shallower observations were used to study the
submm/radio spectral index of the central cluster galaxy (Edge et
al.\ 1999).

Radio counterparts were searched for around the nominal positions of
the submm sources based on the SCUBA astrometry and 2$\sigma$
error-boxes of 6$''$ or 8$''$ diameter depending upon whether the submm
source was respectively a 4$\sigma$ or 3$\sigma$ detection (Fig.~1).
The size of these error-boxes includes both the systematic and random
errors in the source positions and they have been confirmed as
realistic using two sources with CO interferometric observations
(Frayer et al.\ 1998, 1999).  To assign radio fluxes to the individual
submm sources we have adopted the conservative approach of identifying
the radio counterpart as the brightest radio source within the submm
error-box (Fig.~1). The behaviour of the radio-submm spectral index is
such that a brighter radio counterpart leads to a lower redshift
estimate.  Thus by selecting the brightest available radio source, we
should obtain a lower bound on the redshift of the submm source.
Apparent radio fluxes or 3$\sigma$ limits are listed in Table~1.  Note
that a faint radio source lies just outside the error-box of
SMM\,J00265+1710, and so we adopt the radio flux of that source as a
lower limit in this case.  For SMM\,J02399$-$0134, the radio
counterpart is confused due to emission from a nearby bright cluster
elliptical (Fig.~1), and so we only quote an approximate 1.4-GHz flux
for this galaxy.

%
%
\begin{deluxetable}{lccrcccl}
\tablewidth{480pt}
\scriptsize
\tablenum{1}
\label{table-1}
\tablecaption{\sc \small Spectral Properties of 
the Submm Sample\label{tab1}}
\tablehead{
\colhead{Submm} & \colhead{$S_{850}$} & \colhead{$S_{1.4}$} & \colhead{~~$\alpha_{1.4}^{850}$} & \colhead{$z_\alpha$} & \multispan2{~$z_{\rm spec}^d$} &  \colhead{ Comments ~\hfill } \\
\colhead{Source} & \colhead{(mJy)} & \colhead{($\mu$Jy)}  & \colhead{} & \colhead{} & \colhead{Reliable} & \colhead{Uncertain} & \colhead{} }
\startdata
\multispan7{$4\sigma$ Detections  \hfil}\\
\noalign{\smallskip}
SMM\,J02399$-$0136 & 25.4  & 526    & $0.70\pm 0.02$  & 1.1--2.9 & 2.80 & \nodata &  L1/L2 -- Seyfert 2 merger$^a$ \\
SMM\,J00266+1708   & 18.6  & 100    & $0.95\pm 0.04$  & 2.0--4.5 & \nodata & 0.44 &  M8 -- 1$''$ away from radio source\\
SMM\,J09429+4658   & 17.2  & ~32    & $1.14\pm 0.07$  & $>$3.9   & \nodata & \nodata &  H5 -- ERO$^b$ \\
SMM\,J14009+0252   & 14.5  & 529    & $0.60\pm 0.03$  & 0.7--2.3 & \nodata & \nodata &  J5$^c$ -- no spectroscopic observation \\
SMM\,J14011+0252   & 12.3  & 115    & $0.85\pm 0.05$  & 1.7--3.8 & 2.55 & \nodata &  J1/J2 -- starburst merger$^{c}$  \\
SMM\,J02399$-$0134 & 11.0  & $\sim$500 & $0.55\pm 0.04$  & 0.6--2.1 & 1.06 & \nodata &  L3 -- Seyfert 1.5/2 ring galaxy$^{e}$ \\
SMM\,J22471$-$0206 & ~9.2  & $<$65  & $>0.90\pm 0.08$   & $>$1.8 & \nodata & 1.16 & P4 -- weak AGN? \\
SMM\,J02400$-$0134 & ~7.6  & $<$33 & $>0.99\pm 0.11$  & $>$2.4 & \nodata & \nodata &  Optical blank field$^f$    \\
SMM\,J04431+0210   & ~7.2  & $<$70  & $>0.84\pm 0.10$  & $>$1.6 & \nodata & \nodata &  N4 -- ERO$^b$ \\
\noalign{\medskip}
\multispan7{$3\sigma$ Detections \hfil} \\
\noalign{\smallskip}
SMM\,J21536+1742   & ~6.7  & \nodata & \nodata~~~  & \nodata & \nodata & ~1.60? &  No deep radio map \\
SMM\,J00265+1710   & ~6.1  & $\leq$110   & $\geq 0.73\pm 0.07$  & $>$1.2 & \nodata & 0.21 & M6/M10 -- radio ID outside error-box \\
SMM\,J22472$-$0206 & ~6.1  & $<$50 & $>0.87\pm 0.12$  & $>$1.7 & \nodata & ~2.11? &  P2 \\
SMM\,J00266+1710   & ~5.9  & $<$33 & $>0.93\pm 0.12$  & $>$2.9 & \nodata &  0.94 & M3 -- faint radio source at map limit?    \\
SMM\,J00267+1709   & ~5.0  & $<$30 & $>0.92\pm 0.13$  & $>$1.9 &  \nodata & \nodata & Optical blank field$^f$ \\
SMM\,J04433+0210   & ~4.5  & ~70     & $0.75\pm 0.12$  &  1.3--3.2 & \nodata & \nodata &  No spectroscopic observation \\
\noalign{\medskip}
\multispan7{Cluster Galaxies \hfil} \\
\noalign{\smallskip}
SMM\,J21536+1741   & ~9.1  & $226 \times 10^3$ & $-0.58\pm 0.05$  & \nodata & 0.23 &\nodata &  cD -- central galaxy in A\,2390$^{g}$ \\
SMM\,J14010+0252   & ~5.4  & $28.6\times 10^3$  & $-0.30\pm 0.06$ & \nodata & 0.25 &\nodata &  cD -- central galaxy in A\,1835$^{g}$ \\
\enddata
\vspace*{-0.5cm}
\tablerefs{
$a$) Ivison et al.\ (1998) -- 
$b$) Smail et al.\ (1999) -- 
$c$) Ivison et al.\ (1999) --
$d$) Barger et al.\ (1999a) --
$e$) Soucail et al.\ (1999) --
$f$) Smail et al.\ (1998) ---
$g$) Edge et al.\ (1999).
}
\end{deluxetable}

\subsection{Radio-Submm Spectral Indices}

The 850-$\mu$m and 1.4-GHz fluxes or limits for 16 of the sources in
Smail et al.\ (1998), for which we have radio observations, are listed
in Table~1 in order of their apparent submm fluxes, along with their
proposed spectroscopic redshifts, $z_{spec}$, from Barger et
al.\ (1999a). Where a radio counterpart is identified the spectroscopic
redshift of the closest optical candidate is listed in the table.  The
errors on $\alpha_{1.4}^{850}$ are calculated assuming the 1$\sigma$
flux uncertainties in each band (\S2.1 and Smail et al.\ 1998). For
non-detections at 1.4\,GHz we use the 3$\sigma$ flux limit of the
relevant radio map.  The two central cluster galaxies in our
sample are not included in our analysis. 

Using the  $\alpha_{1.4}^{850}$ values or limits, the redshift ranges,
$z_{\alpha}$, are  derived from the extremes of the predictions from
the four CY models (two empirical SEDs representing Arp\,220 and M\,82,
and two models with dust temperatures  of $T_{\rm d}\sim 50$--60\,K and
emissivities of $\beta=1.0$ or $\beta=1.5$) and a further model from
B99, with $\beta=1.5$ and a $T_{\rm d}=30$\,K to illustrate the minimum
possible redshift assuming a very low $T_{\rm d}$.  For a model SED at
$z\gs 1$, a variation of $\delta\beta = +0.2$ or $\delta T_{\rm d} =
-10$\,K results in an change of $\delta \alpha_{1.4}^{850} \ls +0.1$,
equivalent to an uncertainty in the derived redshift of $\delta z/z\sim
0.2$.

There are three caveats to bear in mind when using $\alpha_{1.4}^{850}$
to estimate redshifts for distant galaxies. First, most of the distant
galaxies which CY used to compare with their model predictions show
some signs of AGN activity.  If these AGN also contribute to the
1.4-GHz non-thermal emission of the galaxy they will lower the observed
$\alpha^{850}_{1.4}$ values (as some obviously do in Fig.~1a of CY).
This will mean that any radio-quiet submm sources could lie at the high
end of the predicted $\alpha^{850}_{1.4}$ range at each epoch.
Secondly, the CY and B99 models we use assume effective dust
temperatures for the galaxies, $T_{\rm d}\geq 30$\,K.  If the dust in
distant obscured galaxies is much cooler then this, it will again shift
the predicted redshifts systematically lower (see B99).  Finally, as
mentioned in CY, the effects of inverse Compton scattering of radio
photons off the microwave background may reduce the radio luminosities
of star-forming galaxies at $z\gs 3$ and hence  increase
$\alpha^{850}_{1.4}$ for the most distant galaxies.  Nevertheless, the
relatively good agreement shown in CY of the spectral indices of
distant galaxies with the models is an important confirmation that
$\alpha^{850}_{1.4}$ indices can be used to derive robust lower limits
to the redshifts of submm sources without reliable spectroscopic
identifications.

\section{Results and Discussion}

We show in Fig.~2 three cumulative redshift distributions for the
population  representing extreme interpretations of the $z_\alpha$
limits from the spectral index models.  We see that even making the
most conservative assumptions about the likely redshifts from the
$\alpha^{850}_{1.4}$ indices we still predict a {\it median} redshift
for the submm population above an intrinsic 850-$\mu$m flux of 1\,mJy
of $<\!  z\!>\ \sim 2$, and more likely closer to $z\sim 2.5$--3.

Comparing the cumulative redshift distribution to that derived from the
(incomplete) spectroscopic study of this sample (Barger et al.\ 1999a)
we see broad similarities.  However, comparison of redshifts for
individual sources from the two studies (Table~1), while showing good
agreement for those submm sources with reliable identifications
(e.g.\ Ivison et al.\ 1998, 1999) also indicates that the majority of
the uncertain spectroscopic IDs are likely to be incorrect.   Barger et
al.\ (1999a) obtained spectroscopy of most of the possible optical
counterparts within each submm error-box.  We can therefore state that
the true submm sources must be fainter than the faintest spectroscopic
target.  Including the two optical blank-field sources already known
(Smail et al.\ 1998), we conclude that approximately half of the submm
population are therefore currently unidentified.  These submm sources
have no radio counterparts and are too faint for optical spectroscopy,
$I\gs 24$, their identification will thus be very difficult.

Our median redshift is compatible with the results of Hughes et
al.\ (1998) and CY for the five submm sources in the {\it HDF} based on
analyses of their SEDs and radio-submm indices.  The only other submm
survey for which spectroscopic redshift information has been published
is by Lilly et al.\ (1999) for the Eales et al.\ (1999) sample --- they
find $<\! z\!  >\ \sim 2$ --- similar to our median redshift.  However,
Lilly et al.\ (1999) claim that a third of the submm population lies at
$z<1$; in contrast, we find no galaxies in our field sample at $z<1$.
This apparent contradication may result simply from the small sizes of
the current samples or might indicate that foreground bright optical
galaxies are lensing the distant submm sources detected in the field
surveys (see Blain, M\"oller \& Maller 1999; Hughes et al.\ 1998).

The detection rate of radio counterparts to the submm sources is higher
for the intrinsically brighter sources. All the submm sources with
observed fluxes above $\sim 10$\,mJy (intrinsic fluxes of $\gs 4$\,mJy
) have radio counterparts, while the majority of the fainter sources do
not (this is consistent with CY's results in the {\it HDF}). The
detections and astrometry of the fainter sources are sufficiently
reliable that this result is not due to spurious detections (see Ivison
et al.\ 1999).  Making the conservative assumption of placing all of
the non-detections at their lower bounds on $\alpha_{1.4}^{850}$, we
find that this distribution is consistent with being drawn from the
$\alpha_{1.4}^{850}$ distribution of the brighter sources with a
probability of $P\sim 0.4$.  However, simply comparing the
$\alpha_{1.4}^{850}$ indices for the two subsamples we see that half of
the radio-detected bright submm sources have $\alpha_{1.4}^{850}$
values lower than the lowest limit on the undetected sources, the
likelihood that this occurs by chance is only $P\sim 3\times 10^{-4}$
suggesting that there may be real differences between the
$\alpha_{1.4}^{850}$ values for the two subsamples.  Several factors
could cause this, most simply the apparently fainter submm sources may
be at higher redshifts (a correlation which could naturally exist in a
low density Universe).  Alternatively, intrinsic differences in the
spectral indices of fainter sources would occur if they contain a lower
fraction of radio-loud AGN or have typically cooler dust temperatures
(or higher emissivities).   Both effects are plausible given what we
know about the correlations of AGN fractions, dust temperature and
emissivity with luminosity in local ULIRG samples (Sanders \& Mirabel
1996).  Further detailed observations of both distant and local ULIRGs
are needed to distinguish between these possibilities.

%
%
\hbox{~}
\centerline{\psfig{file=f2.ps,width=3.0in,angle=270}}

\vskip 2mm

\noindent{
\scriptsize \addtolength{\baselineskip}{-3pt} 
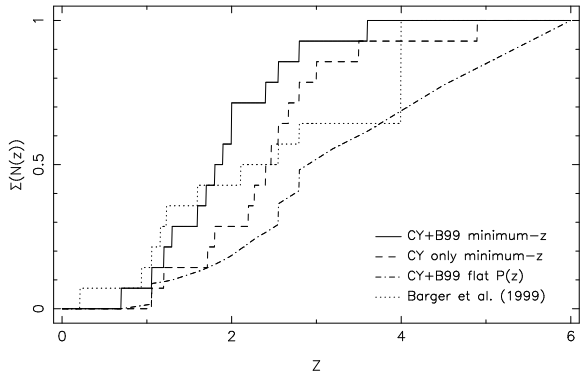
Fig.~2.\ The cumulative redshift distribution for the full submm
sample.   We have used the spectroscopic redshifts of those sources
thought to be reliable (Table~1) and combined these with the probable
redshift ranges of the remaining sources derived from their
$\alpha^{850}_{1.4}$ indices or limits.  The solid line shows the
cumulative distribution if we assume the minimum redshift distribution
which is obtained if {\it all} sources are assumed to lie at their
lower $z_\alpha$ limit given in Table~1 (the dashed line is the
equivalent analysis but restricted to just the CY models).  The effect
of non-thermal radio emission, which drives down the
$\alpha^{850}_{1.4}$ indices, means that this is a very conservative
assumption if some fraction of the population harbor radio-loud AGN.
The dot-dashed line assumes a flat probability distribution for the
sources within their $z_\alpha$ ranges and a maximum redshift of $z=6$
for those sources where we only have a lower limit on
$\alpha^{850}_{1.4}$.  Finally, the dotted line is the cumulative
redshift distribution from Barger et al.\ (1999a) with two of the
source identifications corrected as in Smail et al.\ (1999) and all
blank-field/ERO candidates placed at $z=4$.

\vskip 2mm
\addtolength{\baselineskip}{2pt}
}

The relatively high median redshift we find for the submm population,
$<\!z\! >\ \sim 2$--3, indicates that their equivalent star formation
density at these epochs is around 0.5\,$M_\odot$ yr$^{-1}$ Mpc$^{-3}$
(Blain et al.\ 1999a), roughly three times that seen in UV-selected
samples (Steidel et al.\ 1999).  Emission from dust heated by obscured AGN
will reduce this estimate, but it is difficult not to conclude that
the submm galaxies contain a substantial fraction of the star formation
in the high redshift Universe.

\section{Conclusions}

\noindent{$\bullet$} We present radio maps of 16 galaxies
selected in a deep submm survey.   We combine submm and radio fluxes
(or limits) to determine the radio-submm spectral indices of these
galaxies and interpret these using model predictions to derive
the redshifts for a complete sample of faint submm galaxies.

\noindent{$\bullet$} We find a median redshift $<\!z\! >\ \sim 2$ for
the submm population down to $S_{850}\sim 1$\,mJy under conservative
assumptions, and $<\!z\! >\ \sim 2.5$--3  for more reasonable
assumptions.  Median redshifts below $<\!z\! >\ \ll 2$ are only
possible if the bulk of the emission is coming from dust at $T_{\rm
d}\ll 30$\,K (compared to the 40--50\,K typically seen in well-studied,
distant submm galaxies, or their low-redshift analogs: ULIRGs).  As a
result we find no evidence for a significant low-redshift, $z<1$, tail
in our distribution in contrast to Lilly et al.\ (1999).

\noindent{$\bullet$} We compare the individual redshifts estimated from
$\alpha_{1.4}^{850}$ with the spectroscopic observations of proposed
optical counterparts of the submm sources.  We find that the majority of
the `uncertain' spectroscopic identifications from Barger et al.\ (1999a)
are likely to be incorrect. We conclude that the true counterparts lie
at higher redshifts and are intrinsically very faint, $I\gs 24$, making
the prospects for a complete optical spectroscopic survey of the submm
population bleak.

\section*{Acknowledgements}

We thank Amy Barger, Chris Carilli, Len Cowie, Glenn Morrison, Jason
Stevens and Min Yun for useful conversations and help.   


\end{document}